\documentclass[fleqn]{article}
\usepackage{longtable}
\usepackage{graphicx}

\begin {document}
\begin{flushleft}
{\LARGE
{\bf Comment on ``Relativistic atomic data for W XLVII" by \\S. Aggarwal {\em et al.} [Chin. Phys. B  24 (2015) 053201]}
}\\

\vspace{1.5 cm}

{\bf {Kanti  M.  ~Aggarwal}}\\ 

\vspace*{1.0cm}

Astrophysics Research Centre, School of Mathematics and Physics, Queen's University Belfast, Belfast BT7 1NN, Northern Ireland, UK\\ 
\vspace*{0.5 cm} 

e-mail: K.Aggarwal@qub.ac.uk \\

\vspace*{0.20cm}

Received  20 August 2015.  Accepted for publication 10 October 2015 \\

\vspace*{1.5cm}


\vspace*{1.0 cm}

\hrule

\vspace{0.5 cm}

\end{flushleft}

\clearpage


\begin{abstract}

Recently, S. Aggarwal {\em et al.} [Chin. Phys. B  24 (2015) 053201] reported energy levels, radiative rates and lifetimes for the lowest 148 levels belonging to the 3s$^2$3p, 3s3p$^2$, 3s$^2$3d, 3s3p3d, 3p$^3$, 3p$^2$3d, 3s3d$^2$, 3p3d$^2$, and 3d$^3$ configurations of Al-like tungsten. While their calculated energies for the levels and the radiative rates for transitions are correct, the reported results for lifetimes are completely wrong. According to our calculations, errors in their reported lifetimes are up to 14 orders of magnitude for over 90\% of the levels. Here we report the correct lifetimes and explain the reasons for discrepancies.

\end{abstract}

\clearpage

\section{Introduction}

Tungsten (W), being an important constituent of tokamak reactor walls, is perhaps the most important element for studying  fusion plasmas. It immensely radiates at almost all ionisation stages, and therefore to assess  radiation loss and for modelling plasmas, atomic data (including energy levels and  oscillator strengths or radiative decay rates) are required for many of its ions. The developing  ITER project has raised urgency for the data requirements, and as a result several groups of people are engaged in producing atomic data for its ions. However, the most desirable requirement for (any) atomic data is its accuracy$^{\cite{fst}}$ without which the modelling of plasmas will not be reliable.  Recently, S. Aggarwal {\em et al.}$^{\cite{ajm}}$, henceforth to be referred to as AJM,  have reported results for energy levels, oscillator strengths, radiative rates,  and lifetimes for Al-like W. For their calculations, they adopted the modified version of the {\sc grasp} (general-purpose relativistic atomic structure package) code, available at the website {\tt http://web.am.qub.ac.uk/DARC/}. It is a fully relativistic code,  based on the $jj$ coupling scheme, and hence ideal for generating atomic data for heavier ions, such as of W. Further relativistic corrections arising from the Breit interaction and QED (quantum electrodynamics) effects are also included. 

AJM have reported energies and lifetimes ($\tau$) for the lowest 148 levels of the 3s$^2$3p, 3s3p$^2$, 3s$^2$3d, 3s3p3d, 3p$^3$, 3p$^2$3d, 3s3d$^2$, 3p3d$^2$, and 3d$^3$ (nine) configurations. They have also listed radiative rates (A-values), oscillator strengths (f-values) and  line strengths (S-values) for four types of transitions, namely electric dipole (E1), electric quadrupole (E2), magnetic dipole (M1), and magnetic quadrupole (M2), but only from the ground to higher excited levels. However, unfortunately there are several discrepancies and a few serious errors in their paper, and here we focus only on three, discussed below.

Firstly, their use of the nomenclature of the ion (being wrong) in the title and throughout the text of the paper is confusing and misleading. This is because Al-like tungsten is W LXII, and not W XLVII.   Secondly, to calculate atomic data they have included configuration interaction (CI) among 894 levels of  35 configurations, the additional 26 are: 3s3p4$\ell$, 3s3d4$\ell$, 3p3d4$\ell$, 3s$^2$4$\ell$, 3p$^2$4$\ell$ (except 3p$^2$4d), 3p4$\ell^2$ (except 3p4p$^2$), and 3d4$\ell^2$. However, these 35 configurations generate 1007 levels in total (see section 2), and hence there is a discrepancy of 113 levels. Both of these discrepancies may, at best, be attributed to oversight and do not affect the calculated results. However, the third and the final one is the gross error in their calculated lifetimes, of up to 14 orders of magnitude for over 90\% of the levels -- see section 3. Therefore, the purpose of this comment is to report the correct $\tau$ values and to explain the reasons for these large errors.


\section{Calculations}

For our calculations we adopt the same version of the  {\sc grasp}  code as employed by AJM$^{\cite{ajm}}$. Similarly,  for compatibility we use the same  option of {\em extended average level} (EAL), as adopted by them. To make comparisons, we have performed two sets of calculations, one with 148 levels of 9 configurations (GRASP1), i.e. 3s$^2$3p, 3s3p$^2$, 3s$^2$3d, 3s3p3d, 3p$^3$, 3p$^2$3d, 3s3d$^2$, 3p3d$^2$, and 3d$^3$; and another with 37 (GRASP2), the additional 28  being 3s3p4$\ell$, 3s3d4$\ell$, 3p3d4$\ell$, 3s$^2$4$\ell$, 3p$^2$4$\ell$, 3p4$\ell^2$, and 3d4$\ell^2$. All these configurations, along with the number of levels generated by each, are listed in Table 1. In total, these configurations generate 1056 levels. AJM included the same configurations (except two) in their calculations (GRASP3), but it is not clear why they ignored  3p$^2$4d and 3p4p$^2$. These two configurations (number 28 and 31 in Table 1) generate 49 levels and therefore the number of levels included by AJM should have been 1007 (1056-49), and not 894 as stated by them. However, as far as the calculations of $\tau$ (our parameter of interest) are concerned, the inclusion/exclusion of these two configurations is of no consequence as we will see in the next section.  

The energies obtained in our calculations  for the lowest 148  levels of W LXII are comparable to those reported by AJM in their Table 1, and hence are not repeated here. Similarly, there is no (significant) discrepancy with their A-values, listed in their Table 3. Therefore, we now focus on the $\tau$ values for which we find large discrepancies.

\section{Lifetimes}

The lifetime $\tau$ of a level $j$ is determined as  ${\tau}_j =1/{\sum_{i} A_{ji}}$, i.e. the summation is over all transitions from lower levels $i$ to higher $j$. Generally, A-values for E1 transitions dominate for the determination of $\tau$, but for other type of transitions become important, particularly when the E1 does not exist, such as for 1--3/5/9 -- see Table~2 for level definitions. For this reason,  A-values from all types of transitions are included in a calculation, as was also done by  AJM$^{\cite{ajm}}$. In Table~2, we list our calculated $\tau$ from both GRASP1 and GRASP2 calculations, and also include the results of AJM (GRASP3) for a ready comparison.  

For all 148 levels listed in Table~2, there is no (appreciable) discrepancy between the GRASP1 and GRASP2 results. Therefore, as stated earlier, the inclusion or exclusion of the 3p$^2$4d and 3p4p$^2$ configurations, omitted by AJM, is of no consequence, and hence the  values of $\tau$ calculated by AJM should have been comparable to those of ours. Unfortunately, there are large discrepancies for almost all levels, over 90\% to be precise. For all levels, the reported values of $\tau$ by AJM are invariably {\em higher},  by up to 14 orders of magnitude, see for example levels 11, 12, 25, 26, 37, and 82. Below we explain the possible reasons for these large discrepancies.

For level 3 (3s$^2$3p $^2$P$^o_{3/2}$), the $\tau$ by AJM is higher by a factor of two. For this level, the AJM calculation appears to include only the A-value (2.18$\times$10$^8$ s$^{-1}$) for the 1--3 M1 transition  -- see their Table~3, whereas the A-value for the 2--3 E1 transition (2.64$\times$10$^8$ s$^{-1}$, not listed by AJM) has been completely ignored. Similarly for level 11 (3s3p3d $^4$F$^o_{3/2}$), their calculation is based on the A-value (2.95$\times$10$^6$ s$^{-1}$) of 1--11 M1 transition alone. Interestingly, they have also listed the A-value for the 1--11 E2 transition, i.e. 9.38$\times$10$^6$ s$^{-1}$. The inclusion of these two A-values alone will give $\tau$ = 8.1$\times$10$^{-8}$~s, and {\em not} 3.39$\times$10$^{-7}$~s, as listed by them in their Table~1. However, the correct value for this level is $\sim$2.1$\times$10$^{-12}$~s, as listed in our Table~2. This is because the main contributing transitions are:  2--11 E1 (A=3.72$\times$10$^{11}$ s$^{-1}$), 5--11 E1 (A=5.26$\times$10$^{10}$ s$^{-1}$) and 6--11 E1 (A=3.29$\times$10$^{10}$ s$^{-1}$). Finally, the largest discrepancy of 14 orders of magnitude is for level 82, i.e. 3p3d$^2$ $^4$F$^o_{5/2}$. For this level, AJM have listed A=3.66$\times$10$^{5}$ s$^{-1}$ for the 1--82 E2 transition, which alone gives $\tau$ = 2.73$\times10^{-6}$~s, and {\em not} 5.87$\times$10$^{1}$~s, as listed by them in their Table~1. For this level the dominant contributing transition is 23--82 E1 for which A=2.24$\times$10$^{12}$ s$^{-1}$, giving $\tau$ = 4.46$\times10^{-13}$~s, a value within $\sim$10\% of that listed in present Table~2. Therefore, it is abundantly clear that the $\tau$ values reported by AJM$^{\cite{ajm}}$ are simply wrong for most of the levels. 

\section{Conclusions}

In this work, we have calculated energy levels, radiative rates and lifetimes for the lowest 148 levels/transitions of Al-like tungsten, i.e. W LXII. For the calculations, the well known {\sc grasp} code has been adopted, as by S. Aggarwal {\em et al.}$^{\cite{ajm}}$ who have recently reported similar results. However, their results for energy levels and A-values (although for limited transitions) are correct, the corresponding values of $\tau$ are in huge errors, of up to 14 orders of magnitude, for almost all levels. In fact, their reported $\tau$ appear to have no relationship with the A-values for transitions of this ion. Therefore, for the benefit of users as well as for future workers, we have listed the correct values of $\tau$, and have also explained the reasons for discrepancies. Apart from the errors in $\tau$, other discrepancies in the paper of AJM have been noted. However, we stress that the listed anomalies are not the only ones, there are a few more. As an example, the 101--129 levels listed in their Table~1 have no (odd) parity indication.

Finally, the A-values listed by S. Aggarwal {\em et al.}$^{\cite{ajm}}$ for transitions from the ground level alone are not sufficient for applications, because a {\em complete} set of data for {\em all} transitions are required for any modelling application. Besides this, there is scope for improvement in their work, because in their CI calculations they have ignored the inclusion of some of the important configurations, such as 3p$^2$4d, 3p4p$^2$, 3d$^2$4$\ell$,  and 3$\ell$4$\ell\ell'$, apart from those of $n$ = 5. Therefore, an improved set of complete data for all transitions of W LXII are reported in a separate paper$^{\cite{w62}}$.



\newpage
\clearpage

\begin{flushleft}
Table 1. Configurations and  levels of W LXII. Odd parity levels are indicated by a superscript `o'.
\end{flushleft}
\begin{tabular}{rllrrrrrrrrr} \hline
\\
Index  & Configuration      & No. of Levels  & Total No. of Levels  \\
\\ \hline
  1  &  3s$^2$3p   &    2$^o$   &    2  \\
  2  &  3s3p$^2$   &    8       &   10  \\
  3  &  3s$^2$3d   &    2       &   12  \\
  4  &  3s3p3d	   &   23$^o$   &   35  \\
  5  &  3p$^3$	   &    5$^o$   &   40  \\
  6  &  3p$^2$3d   &   28       &   68  \\
  7  &  3s3d$^2$   &   16       &   84  \\
  8  &  3p3d$^2$   &   45$^o$   &  129  \\
  9  &  3d$^3$	   &   19       &  148  \\
 10  &  3s3p4s	   &    7$^o$   &  155  \\
 11  &  3s3p4p	   &   18       &  173  \\
 12  &  3s3p4d	   &   23$^o$   &  196  \\
 13  &  3s3p4f	   &   24       &  220  \\
 14  &  3s3d4s	   &    8       &  228  \\
 15  &  3s3d4p	   &   23$^o$   &  251  \\
 16  &  3s3d4d	   &   34       &  285  \\
 17  &  3s3d4f	   &   39$^o$   &  324  \\
 18  &  3p3d4s	   &   23$^o$   &  347  \\
 19  &  3p3d4p	   &   65       &  412  \\
 20  &  3p3d4d	   &   96$^o$   &  508  \\
 21  &  3p3d4f	   &  113       &  621  \\
 22  &  3s$^2$4s   &    1       &  622  \\
 23  &  3s$^2$4p   &    2$^o$   &  624  \\
 24  &  3s$^2$4d   &    2       &  626  \\
 25  &  3s$^2$4f   &    2$^o$   &  628  \\
 26  &  3p$^2$4s   &    8       &  636  \\
 27  &  3p$^2$4p   &   21$^o$   &  657  \\
 28  &  3p$^2$4d   &   28       &  685  \\
 29  &  3p$^2$4f   &   30$^o$   &  715  \\
 30  &  3p4s$^2$   &    2$^o$   &  717  \\
 31  &  3p4p$^2$   &   21$^o$   &  738  \\
 32  &  3p4d$^2$   &   45$^o$   &  783  \\
 33  &  3p4f$^2$   &   69$^o$   &  852  \\
 34  &  3d4s$^2$   &    2       &  854  \\
 35  &  3d4p$^2$   &   28       &  882  \\
 36  &  3d4d$^2$   &   67       &  949  \\
 37  &  3d4f$^2$   &  107       & 1056  \\
 \\ \hline  											      
\end{tabular}   									   					       
			      							   					       
\vspace*{0.5 cm}													       
\begin{flushleft}													       
{\small
									       
}															       
\end{flushleft} 

\newpage
\clearpage
\begin{flushleft}
Table 2. Lifetimes ($\tau$, s) for the  levels of W LXII. $a{\pm}b \equiv a{\times}$10$^{{\pm}b}$.
\end{flushleft}
{\small
\begin{tabular}{rllcrrrrrrrrr} \hline
\\
Index  & Configuration      & Level  & GRASP1 & GRASP2  & GRASP3 \\
\\ \hline
1   &	3s$^2$3p          &  $^2$P$^o_{1/2}$  &  & & .........  \\    
2   &	3s3p$^2$($^3$P)   &  $^4$P$  _{1/2}$  & 2.550$-$11 &  2.538$-$11 & 2.54$-$11  \\	 
3   &	3s$^2$3p          &  $^2$P$^o_{3/2}$  & 2.071$-$09 &  2.046$-$09 & 4.58$-$09  \\    
4   &	3s3p$^2$($^3$P)   &  $^4$P$  _{3/2}$  & 7.446$-$11 &  7.357$-$11 & 8.85$-$11  \\	  
5   &	3s3p$^2$($^1$D)   &  $^2$D$  _{5/2}$  & 7.026$-$11 &  6.988$-$11 & 6.99$-$11  \\	  
6   &	3s3p$^2$($^1$D)   &  $^2$D$  _{3/2}$  & 1.448$-$12 &  1.443$-$12 & 1.51$-$12  \\	  
7   &	3s3p$^2$($^3$P)   &  $^2$P$  _{1/2}$  & 3.493$-$13 &  3.490$-$13 & 3.53$-$13  \\	  
8   &	3s$^2$3d          &  $^2$D$  _{3/2}$  & 2.577$-$13 &  2.577$-$13 & 2.58$-$13  \\    
9   &	3s$^2$3d          &  $^2$D$  _{5/2}$  & 3.417$-$12 &  3.425$-$12 & 3.43$-$12  \\    
10  &	3p$^3$            &  $^2$P$^o_{3/2}$  & 4.586$-$12 &  4.601$-$12 & 6.61$-$07  \\    
11  &	3s3p($^3$P)3d     &  $^4$F$^o_{3/2}$  & 2.148$-$12 &  2.159$-$12 & 3.39$-$07  \\       
12  &	3s3p($^3$P)3d     &  $^4$F$^o_{5/2}$  & 3.473$-$11 &  3.473$-$11 & 9.26$-$04  \\       
13  &	3s3p($^3$P)3d     &  $^4$D$^o_{1/2}$  & 4.189$-$13 &  4.195$-$13 & 8.51$-$06  \\       
14  &	3s3p($^3$P)3d     &  $^4$D$^o_{3/2}$  & 2.863$-$13 &  2.858$-$13 & 4.41$-$05  \\       
15  &	3s3p($^3$P)3d     &  $^4$P$^o_{5/2}$  & 3.519$-$12 &  3.530$-$12 & 2.69$-$06  \\       
16  &	3s3p($^3$P)3d     &  $^4$F$^o_{7/2}$  & 4.710$-$12 &  4.715$-$12 & 5.34$-$12  \\       
17  &	3s3p($^3$P)3d     &  $^2$F$^o_{5/2}$  & 3.871$-$12 &  3.870$-$12 & 8.53$-$06  \\       
18  &	3s3p($^3$P)3d     &  $^2$D$^o_{3/2}$  & 1.401$-$12 &  1.405$-$12 & 5.22$-$06  \\       
19  &	3s3p$^2$($^3$P)   &  $^4$P$  _{5/2}$  & 1.397$-$12 &  1.399$-$12 & 1.40$-$12  \\	  
20  &	3s3p$^2$($^3$P)   &  $^2$S$  _{1/2}$  & 4.880$-$13 &  4.869$-$13 & 1.32$-$10  \\	  
21  &	3p$^2$3d          &  $^4$F$^o_{3/2}$  & 1.542$-$12 &  1.518$-$12 & 3.50$-$08  \\    
22  &	3s3p$^2$          &  $^2$P$  _{3/2}$  & 2.729$-$13 &  2.739$-$13 & 4.35$-$10  \\    
23  &	3p$^2$3d          &  $^4$F$  _{5/2}$  & 3.051$-$12 &  3.066$-$12 & 9.39$-$09  \\    
24  &	3p$^3$($^2$D)     &  $^2$D$^o_{5/2}$  & 8.241$-$13 &  8.260$-$13 & 6.38$-$07  \\	    
25  &	3p$^3$($^4$S)     &  $^4$S$^o_{3/2}$  & 3.188$-$13 &  3.195$-$13 & 1.33$-$04  \\	    
26  &	3p$^3$($^2$P)     &  $^2$P$^o_{1/2}$  & 4.161$-$13 &  4.163$-$13 & 1.52$-$05  \\	    
27  &	3s3p($^3$P)3d     &  $^4$P$^o_{3/2}$  & 4.181$-$13 &  4.190$-$13 & 1.40$-$06  \\       
28  &	3s3p($^3$P)3d     &  $^4$P$^o_{1/2}$  & 5.586$-$13 &  5.596$-$13 & 1.24$-$06  \\       
29  &	3s3p($^3$P)3d     &  $^4$D$^o_{5/2}$  & 3.170$-$13 &  3.177$-$13 & 5.19$-$04  \\       
30  &	3s3p($^3$P)3d     &  $^4$D$^o_{7/2}$  & 5.970$-$13 &  5.981$-$13 & 6.11$-$13  \\       
31  &	3s3p($^3$P)3d     &  $^2$D$^o_{3/2}$  & 2.275$-$13 &  2.277$-$13 & 1.19$-$05  \\       
32  &	3s3p($^3$P)3d     &  $^2$F$^o_{5/2}$  & 1.948$-$13 &  1.961$-$13 & 1.36$-$03  \\       
33  &	3s3p($^3$P)3d     &  $^2$P$^o_{1/2}$  & 1.746$-$13 &  1.757$-$13 & 3.82$-$05  \\       
34  &	3s3p($^3$P)3d     &  $^4$F$^o_{9/2}$  & 4.707$-$09 &  4.702$-$09 & 4.85$-$09  \\       
35  &	3s3p($^3$P)3d     &  $^2$D$^o_{5/2}$  & 1.196$-$12 &  1.206$-$12 & 7.47$-$06  \\       
36  &	3s3p($^3$P)3d     &  $^2$F$^o_{7/2}$  & 1.826$-$12 &  1.833$-$12 & 9.17$-$12  \\ 
37  &	3s3p($^3$P)3d     &  $^2$P$^o_{3/2}$  & 1.109$-$12 &  1.123$-$12 & 1.19$-$04  \\           
 \hline  											      
\end{tabular} 
\newpage
\begin{tabular}{rllcrrrrrrrrr} \hline
\\
Index  & Configuration      & Level  & GRASP1 & GRASP2  & GRASP3 \\
\\ \hline       
38  &	3s3p($^1$P)3d     &  $^2$F$^o_{7/2}$  & 4.036$-$13 &  4.087$-$13 & 3.27$-$11  \\       
39  &	3s3p($^3$P)3d     &  $^2$P$^o_{1/2}$  & 1.243$-$12 &  1.235$-$12 & 1.28$-$03  \\       
40  &	3s3p($^1$P)3d     &  $^2$P$^o_{3/2}$  & 3.305$-$13 &  3.345$-$13 & 1.59$-$06  \\       
41  &	3s3p($^1$P)3d     &  $^2$D$^o_{5/2}$  & 3.334$-$13 &  3.346$-$13 & 5.50$-$06  \\       
42  &	3p$^2$($^1$D)3d   &  $^2$F$  _{5/2}$  & 9.155$-$13 &  9.226$-$13 & 1.75$-$08  \\	  
43  &	3p$^2$($^3$P)3d   &  $^4$D$  _{3/2}$  & 5.132$-$13 &  5.172$-$13 & 2.47$-$09  \\	  
44  &	3p$^2$($^3$P)3d   &  $^4$D$  _{1/2}$  & 4.601$-$13 &  4.648$-$13 & 4.57$-$09  \\	  
45  &	3p$^2$($^1$D)3d   &  $^2$G$  _{7/2}$  & 6.025$-$13 &  6.109$-$13 & 2.30$-$05  \\	  
46  &	3p$^2$($^1$D)3d   &  $^2$D$  _{3/2}$  & 2.936$-$13 &  2.941$-$13 & 2.73$-$09  \\	  
47  &	3p$^2$($^3$P)3d   &  $^2$D$  _{5/2}$  & 3.316$-$13 &  3.330$-$13 & 8.89$-$09  \\	  
48  &	3p$^2$($^3$P)3d   &  $^4$D$  _{1/2}$  & 3.681$-$13 &  3.674$-$13 & 2.58$-$07  \\	  
49  &	3s3d$^2$($^3$F)   &  $^4$F$  _{3/2}$  & 2.333$-$13 &  2.348$-$13 & 1.46$-$06  \\	  
50  &	3s3d$^2$($^3$F)   &  $^4$F$  _{5/2}$  & 2.061$-$13 &  2.072$-$13 & 1.03$-$09  \\  
51  &	3p$^2$($^3$P)3d    &  $^4$F$  _{7/2}$  & 6.144$-$13 &  6.203$-$13 & 2.36$-$05  \\	 
52  &	3p$^2$($^1$D)3d    &  $^2$G$  _{9/2}$  & 1.177$-$12 &  1.183$-$12 & 1.29$-$12  \\	 
53  &	3s3d$^2$($^3$P)    &  $^4$P$  _{1/2}$  & 1.950$-$13 &  1.979$-$13 & 1.75$-$08  \\	 
54  &	3p$^2$($^1$D)3d    &  $^2$P$  _{3/2}$  & 6.222$-$13 &  6.253$-$13 & 3.12$-$08  \\	 
55  &	3p$^2$($^1$D)3d    &  $^2$D$  _{5/2}$  & 6.468$-$13 &  6.516$-$13 & 1.35$-$08  \\	 
56  &	3p$^2$($^1$D)3d    &  $^2$F$  _{7/2}$  & 9.228$-$13 &  9.346$-$13 & 2.49$-$05  \\	 
57  &	3p$^2$($^3$P)3d    &  $^4$P$  _{5/2}$  & 4.992$-$13 &  5.063$-$13 & 2.44$-$07  \\	 
58  &	3p$^2$($^1$D)3d    &  $^2$S$  _{1/2}$  & 8.612$-$13 &  8.652$-$13 & 1.46$-$10  \\	 
59  &	3p$^2$($^3$P)3d    &  $^2$D$  _{3/2}$  & 5.006$-$13 &  5.067$-$13 & 1.06$-$10  \\	 
60  &	3s3d$^2$($^3$F)    &  $^4$F$  _{7/2}$  & 3.663$-$13 &  3.693$-$13 & 6.83$-$02  \\	 
61  &	3s3d$^2$($^3$P)    &  $^4$P$  _{5/2}$  & 4.248$-$13 &  4.292$-$13 & 1.10$-$08  \\	 
62  &	3s3d$^2$($^3$P)    &  $^4$P$  _{3/2}$  & 4.498$-$13 &  4.552$-$13 & 2.72$-$09  \\	 
63  &	3s3d$^2$($^1$G)    &  $^2$G$  _{9/2}$  & 3.807$-$13 &  3.873$-$13 & 4.49$-$13  \\	 
64  &	3s3d$^2$($^3$F)    &  $^2$F$  _{5/2}$  & 2.996$-$13 &  3.023$-$13 & 1.75$-$08  \\	 
65  &	3s3d$^2$($^1$G)    &  $^2$G$  _{7/2}$  & 4.601$-$13 &  4.678$-$13 & 3.48$-$04  \\	 
66  &	3s3d$^2$($^1$D)    &  $^2$D$  _{3/2}$  & 3.101$-$13 &  3.141$-$13 & 1.02$-$10  \\	 
67  &	3s3d$^2$($^3$P)    &  $^2$P$  _{1/2}$  & 2.577$-$13 &  2.619$-$13 & 7.30$-$11  \\	 
68  &	3p$^3$($^2$D)      &  $^2$D$^o_{3/2}$  & 1.695$-$13 &  1.698$-$13 & 1.33$-$03  \\ 
69  &	3p3d$^2$($^1$G)    &  $^2$F$^o_{5/2}$  & 1.782$-$12 &  1.812$-$12 & 3.35$-$02  \\	 
70  &	3s3d$^2$($^3$F)    &  $^4$F$  _{9/2}$  & 1.447$-$12 &  1.475$-$12 & 7.27$-$12  \\	 
71  &	3s3d$^2$($^1$D)    &  $^2$D$  _{5/2}$  & 1.232$-$12 &  1.255$-$12 & 3.15$-$10  \\	 
72  &	3p3d$^2$($^3$P)    &  $^4$D$^o_{1/2}$  & 5.100$-$13 &  5.165$-$13 & 7.47$-$03  \\	 
73  &	3p3d$^2$($^3$F)    &  $^2$D$^o_{3/2}$  & 2.800$-$13 &  2.799$-$13 & 1.11$-$03  \\	 
74  &	3s3d$^2$($^3$F)    &  $^2$F$  _{7/2}$  & 1.194$-$12 &  1.213$-$12 & 9.25$-$03  \\	 
\hline  											      
\end{tabular} 
\newpage
\begin{tabular}{rllcrrrrrrrrr} \hline
\\
Index  & Configuration      & Level  & GRASP1 & GRASP2  & GRASP3 \\
\\ \hline   
75  &	3s3d$^2$($^3$P)    &  $^2$P$  _{3/2}$  & 9.842$-$13 &  1.001$-$12 & 2.61$-$08  \\	 
76  &	3s3d$^2$($^1$S)    &  $^2$S$  _{1/2}$  & 9.985$-$13 &  1.024$-$12 & 1.81$-$09  \\	 
77  &	3p3d$^2$($^3$F)    &  $^4$G$^o_{7/2}$  & 3.497$-$12 &  3.521$-$12 & 2.21$-$09  \\	 
78  &	3p3d$^2$($^3$F)    &  $^4$G$^o_{5/2}$  & 2.370$-$12 &  2.402$-$12 & 5.79$-$03  \\	 
79  &	3p3d$^2$($^3$P)    &  $^4$D$^o_{3/2}$  & 2.426$-$12 &  2.454$-$12 & 3.42$-$04  \\	 
80  &	3p3d$^2$($^3$P)    &  $^4$P$^o_{1/2}$  & 2.505$-$12 &  2.531$-$12 & 7.52$-$05  \\	 
81  &	3p3d$^2$($^1$G)    &  $^2$H$^o_{9/2}$  & 3.533$-$12 &  3.570$-$12 & 8.28$-$06  \\	 
82  &	3p3d$^2$($^3$F)    &  $^4$F$^o_{5/2}$  & 3.968$-$13 &  3.965$-$13 & 5.87$+$01  \\	 
83  &	3p3d$^2$($^1$G)    &  $^2$F$^o_{7/2}$  & 4.454$-$13 &  4.522$-$13 & 3.28$-$08  \\	 
84  &	3p3d$^2$($^3$P)    &  $^4$S$^o_{3/2}$  & 3.570$-$13 &  3.611$-$13 & 5.11$-$05  \\	 
85  &	3p$^2$3d($^2$D)    &  $^4$P$  _{3/2}$  & 1.778$-$13 &  1.789$-$13 & 3.50$-$08  \\	 
86  &	3p$^2$3d($^2$D)    &  $^4$D$  _{5/2}$  & 1.847$-$13 &  1.864$-$13 & 1.27$-$08  \\	 
87  &	3p$^2$3d($^2$D)    &  $^2$P$  _{1/2}$  & 1.745$-$13 &  1.759$-$13 & 4.77$-$09  \\	 
88  &	3p$^2$3d($^2$D)    &  $^2$F$  _{7/2}$  & 1.926$-$13 &  1.949$-$13 & 4.33$-$06  \\	 
89  &	3p3d$^2$($^3$F)    &  $^4$D$^o_{7/2}$  & 1.715$-$12 &  1.735$-$12 & 2.75$-$08  \\	 
90  &	3p3d$^2$($^3$F)    &  $^2$G$^o_{9/2}$  & 2.438$-$12 &  2.450$-$12 & 3.83$-$01  \\	 
91  &	3p3d$^2$($^1$D)    &  $^2$P$^o_{3/2}$  & 1.571$-$12 &  1.585$-$12 & 5.55$-$03  \\	 
92  &	3p3d$^2$($^3$P)    &  $^4$D$^o_{5/2}$  & 1.876$-$12 &  1.890$-$12 & 1.23$-$03  \\	 
93  &	3p$^2$3d($^2$D)    &  $^2$D$  _{3/2}$  & 1.882$-$13 &  1.900$-$13 & 2.85$-$09  \\	 
94  &	3p$^2$3d($^2$D)    &  $^4$D$  _{1/2}$  & 1.342$-$12 &  1.375$-$12 & 1.98$-$01  \\	 
95  &	3p$^2$3d($^2$D)    &  $^4$F$  _{9/2}$  & 2.808$-$13 &  2.842$-$13 & 2.75$-$09  \\	 
96  &	3p$^2$3d($^2$D)    &  $^4$D$  _{7/2}$  & 2.637$-$13 &  2.668$-$13 & 6.13$-$05  \\	 
97  &	3p$^2$3d($^2$D)    &  $^2$P$  _{1/2}$  & 2.369$-$13 &  2.395$-$13 & 1.48$-$07  \\	 
98  &	3p$^2$3d($^2$D)    &  $^2$D$  _{5/2}$  & 2.621$-$13 &  2.653$-$13 & 1.49$-$10  \\	 
99  &	3p$^2$3d($^2$D)    &  $^2$P$  _{3/2}$  & 2.244$-$13 &  2.275$-$13 & 9.46$-$04  \\	 
100 &	3p$^2$3d($^2$D)    &  $^2$F$  _{5/2}$  & 2.305$-$13 &  2.334$-$13 & 3.18$-$10  \\      
101 &	3p3d$^2$($^3$F)    &  $^4$F$^o_{3/2}$  & 1.876$-$13 &  1.894$-$13 & 4.84$-$02  \\	 
102 &	3p3d$^2$($^3$F)    &  $^2$F$^o_{5/2}$  & 1.979$-$13 &  2.001$-$13 & 2.73$-$03  \\	 
103 &	3p3d$^2$($^3$F)    &  $^4$D$^o_{1/2}$  & 1.857$-$13 &  1.874$-$13 & 1.85$-$02  \\	 
104 &	3p3d$^2$($^3$F)    &  $^2$G$^o_{7/2}$  & 2.060$-$13 &  2.084$-$13 & 4.13$-$09  \\	 
105 &	3p3d$^2$($^3$P)    &  $^4$P$^o_{3/2}$  & 1.871$-$13 &  1.902$-$13 & 2.73$-$03  \\	 
106 &	3p3d$^2$($^3$F)    &  $^4$G$^o_{9/2}$  & 3.026$-$13 &  3.069$-$13 & 3.20$-$06  \\	 
107 &	3p3d$^2$($^3$P)    &  $^4$D$^o_{7/2}$  & 2.832$-$13 &  2.870$-$13 & 3.31$-$10  \\	 
108 &	3p3d$^2$($^3$P)    &  $^4$P$^o_{5/2}$  & 2.753$-$13 &  2.798$-$13 & 5.42$-$05  \\	 
109 &	3p3d$^2$($^1$G)    &  $^2$H$^o_{11/2}$ & 3.523$-$13 &  3.600$-$13 & 2.94$-$04  \\	 
110 &	3p3d$^2$($^3$P)    &  $^2$S$^o_{1/2}$  & 2.568$-$13 &  2.600$-$13 & 1.71$-$02  \\	 
111 &	3p3d$^2$($^3$F)    &  $^4$D$^o_{3/2}$  & 2.543$-$13 &  2.577$-$13 & 9.28$-$03  \\	 
\hline  											      
\end{tabular} 
\newpage
\begin{tabular}{rllcrrrrrrrrr} \hline
\\
Index  & Configuration      & Level  & GRASP1 & GRASP2  & GRASP3 \\
\\ \hline
112 &	3p3d$^2$($^3$F)    &  $^4$D$^o_{5/2}$  & 2.638$-$13 &  2.675$-$13 & 2.02$-$01  \\	 
113 &	3p3d$^2$($^1$G)    &  $^2$G$^o_{7/2}$  & 2.888$-$13 &  2.938$-$13 & 6.64$-$10  \\	 
114 &	3p3d$^2$($^3$F)    &  $^4$F$^o_{7/2}$  & 2.668$-$13 &  2.711$-$13 & 1.81$-$09  \\	 
115 &	3p3d$^2$($^3$P)    &  $^2$D$^o_{5/2}$  & 2.520$-$13 &  2.559$-$13 & 2.37$-$04  \\	 
116 &	3p3d$^2$($^1$D)    &  $^2$D$^o_{3/2}$  & 2.444$-$13 &  2.476$-$13 & 1.43$-$02  \\	 
117 &	3p3d$^2$($^1$G)    &  $^2$G$^o_{9/2}$  & 2.828$-$13 &  2.883$-$13 & 1.03$-$05  \\	 
118 &	3p3d$^2$($^3$P)    &  $^2$D$^o_{3/2}$  & 2.469$-$13 &  2.513$-$13 & 2.04$-$01  \\	 
119 &	3p3d$^2$($^1$G)    &  $^2$F$^o_{5/2}$  & 2.591$-$13 &  2.642$-$13 & 5.34$-$04  \\	 
120 &	3p3d$^2$($^3$P)    &  $^2$P$^o_{1/2}$  & 2.444$-$13 &  2.485$-$13 & 3.23$-$02  \\	 
121 &	3p3d$^2$($^3$F)    &  $^4$G$^o_{11/2}$ & 5.601$-$13 &  5.737$-$13 & 1.81$-$05  \\	 
122 &	3p3d$^2$($^1$D)    &  $^2$D$^o_{5/2}$  & 4.432$-$13 &  4.524$-$13 & 3.29$-$03  \\	 
123 &	3p3d$^2$($^1$D)    &  $^2$F$^o_{7/2}$  & 4.900$-$13 &  5.005$-$13 & 9.14$-$10  \\	 
124 &	3p3d$^2$($^3$F)    &  $^4$F$^o_{9/2}$  & 4.476$-$13 &  4.555$-$13 & 5.21$-$05  \\	 
125 &	3p3d$^2$($^1$D)    &  $^2$P$^o_{1/2}$  & 3.904$-$13 &  3.975$-$13 & 8.02$-$00  \\	 
126 &	3p3d$^2$($^3$F)    &  $^2$F$^o_{7/2}$  & 3.816$-$13 &  3.895$-$13 & 6.02$-$09  \\	 
127 &	3p3d$^2$($^1$S)    &  $^2$P$^o_{3/2}$  & 4.347$-$13 &  4.459$-$13 & 7.53$-$03  \\	 
128 &	3p3d$^2$($^3$F)    &  $^2$D$^o_{5/2}$  & 3.792$-$13 &  3.872$-$13 & 3.51$-$03  \\	 
129 &	3p3d$^2$($^3$P)    &  $^2$P$^o_{3/2}$  & 3.470$-$13 &  3.546$-$13 & 1.11$-$02  \\	 
130 &	3d$^3$($^4$F)      &  $^4$F$  _{3/2}$  & 1.940$-$13 &  1.952$-$13 & 6.38$-$07  \\ 
131 &	3d$^3$($^4$F)      &  $^4$F$  _{5/2}$  & 2.735$-$13 &  2.755$-$13 & 1.08$-$06  \\ 
132 &	3d$^3$($^4$P)      &  $^4$P$  _{3/2}$  & 2.626$-$13 &  2.641$-$13 & 5.31$-$05  \\ 
133 &	3d$^3$($^2$H)      &  $^2$H$  _{9/2}$  & 3.068$-$13 &  3.128$-$13 & 1.96$-$09  \\ 
134 &	3d$^3$($^2$G)      &  $^2$G$  _{7/2}$  & 2.896$-$13 &  2.935$-$13 & 6.02$-$03  \\ 
135 &	3d$^3$($^2$D)      &  $^4$P$  _{1/2}$  & 2.584$-$13 &  2.594$-$13 & 5.77$-$06  \\ 
136 &	3d$^3$($^4$P)      &  $^2$D$  _{5/2}$  & 2.631$-$13 &  2.655$-$13 & 8.65$-$06  \\ 
137 &	3d$^3$($^4$F)      &  $^4$F$  _{7/2}$  & 4.503$-$13 &  4.544$-$13 & 5.76$-$03  \\ 
138 &	3d$^3$($^4$F)      &  $^4$F$  _{9/2}$  & 4.904$-$13 &  4.983$-$13 & 4.82$-$10  \\ 
139 &	3d$^3$($^2$D)      &  $^2$D$  _{3/2}$  & 4.274$-$13 &  4.305$-$13 & 1.62$-$06  \\ 
140 &	3d$^3$($^2$P)      &  $^2$P$  _{1/2}$  & 4.102$-$13 &  4.117$-$13 & 6.84$-$05  \\ 
141 &	3d$^3$($^2$H)      &  $^2$H$  _{11/2}$ & 5.645$-$13 &  5.832$-$13 & 1.17$-$05  \\ 
142 &	3d$^3$($^4$P)      &  $^4$P$  _{5/2}$  & 3.976$-$13 &  3.994$-$13 & 8.09$-$09  \\ 
143 &	3d$^3$($^2$D)      &  $^2$D$  _{5/2}$  & 4.565$-$13 &  4.641$-$13 & 8.94$-$06  \\ 
144 &	3d$^3$($^2$F)      &  $^2$F$  _{7/2}$  & 4.730$-$13 &  4.831$-$13 & 1.27$-$02  \\ 
145 &	3d$^3$($^2$D)      &  $^2$D$  _{3/2}$  & 4.373$-$13 &  4.436$-$13 & 1.35$-$06  \\ 
146 &	3d$^3$($^2$G)      &  $^2$G$  _{9/2}$  & 1.286$-$12 &  1.302$-$12 & 1.39$-$09  \\ 
147 &	3d$^3$($^4$P)      &  $^2$P$  _{3/2}$  & 1.031$-$12 &  1.041$-$12 & 2.43$-$05  \\ 
148 &	3d$^3$($^2$F)      &  $^2$F$  _{5/2}$  & 9.786$-$13 &  9.848$-$13 & 5.25$-$07  \\												      
 \hline  						    					      
\end{tabular}   					    								       
			      							   					       
\begin{flushleft}													       
{\small
GRASP1: present calculations with the {\sc grasp} code with 148 levels \\
GRASP2: present calculations with the {\sc grasp} code with 1056 levels \\
GRASP3: earlier calculations of S. Aggarwal {\em et al.}$^{\cite{ajm}}$ ~with the {\sc grasp} code with 894 levels \\									       
}															       
\end{flushleft} 
}													       
\end{document}